\documentclass[12pt,preprint,showpacs,superscriptaddress,pre]{revtex4}
\usepackage{amsmath}
\usepackage{amssymb}
\usepackage[usenames]{color}
\usepackage[dvips]{graphicx}

\begin{document}

\title{Distribution of averages in a correlated Gaussian medium as a tool for the estimation of the cluster distribution on size}

\author{S.V. Novikov}
\affiliation{Department of Chemistry and Institute of Nanoscale Physics and Chemistry, Katholieke Universiteit Leuven, Celestijnenlaan 200F, B-3001 Leuven, Belgium}
\affiliation{A.N. Frumkin Institute of Physical Chemistry and Electrochemistry, Leninsky prosp. 31, 119991 Moscow, Russia}
\author{M. Van der Auweraer}
\affiliation{Department of Chemistry and Institute of Nanoscale Physics and Chemistry, Katholieke Universiteit Leuven, Celestijnenlaan 200F, B-3001 Leuven, Belgium}


\begin{abstract}
Calculation of the distribution of the average value of a Gaussian random field in a finite domain is carried out for different cases. The results of the calculation demonstrate a strong dependence of the width of the distribution on the spatial correlations of the field. Comparison with the simulation results for the distribution of the size of the cluster indicates that the distribution of an average field could serve as a useful tool for the estimation of the asymptotic behavior of the distribution of the size of the clusters for "deep" clusters where value of the field on each site is much greater than the rms disorder.
\end{abstract}

\maketitle

\section{INTRODUCTION}

A common feature of any random medium is the formation of clusters. Probably, the most well-known problem where statistics of clusters has been extensively studied is the famous percolation problem \cite{Broadbent:629}. In the lattice percolation model any site is occupied with the probability $p$ and non-occupied with the probability $1-p$, and each cluster is a set of connected occupied sites. An important characteristic of the random medium is distribution of the size of the clusters, or, more exactly, the number of clusters with $s$ sites per lattice site,  $n_s$. Knowledge of the statistics of random clusters is vital for the description of many important natural phenomena, such as the conductivity of disordered materials, the flow of liquids in porous media, fracture processes in materials, or even the dynamics of landscapes and forest fires \cite{Shklovskii:book,Turner:171,Reed:239,Drossel:1629,Sahimi:213}. Exact or reliable approximate analytic results for $n_s$ are not very numerous. Possible examples include the number of percolation clusters for $s\gg 1$ and $p\rightarrow 0$
\begin{equation}
n_s \propto s^{-\theta}p^{-s},
\label{perc_deep}
\end{equation}
or the corresponding distribution near the percolation threshold $p_c$
\begin{equation}
n_s \propto s^{-\tau}\exp\left[C(p-p_c)s^\delta\right], \hskip10pt p\rightarrow p_c, \hskip10pt s\gg 1.
\label{perc_near}
\end{equation}
Here $\theta$, $\tau$, $C$, and $\delta$ are some constants \cite{Parisi:871,Stauffer:book,Kunz:77}. Most results in this area were obtained  using scaling arguments with subsequent testing of their validity  with extensive computer simulation \cite{Stauffer:book,Rapaport:679,Grassberger:036101}. The reason for the scarcity of analytical results is obvious: it is difficult to take into account various shapes of clusters. In addition, most known results for $n_s$ are obtained for the uncorrelated case, i.e. the case when lattice sites are occupied independently of each other.

Cluster numbers for the case, where sites are occupied not independently, have been studied for the problem of correlated percolation \cite{Frary:4323,Coniglio:1773,Tuthill:6389,Moura:1023,Nakanishi:693}. Correlation is usually introduced by the short range interaction between different sites (most popular cases are the Ising model \cite{Coniglio:205,Domb:1141} and the q-states Potts model \cite{Jan:L699,Hu:6491}). Attention in this area was almost exclusively focused on the behavior in the vicinity of the percolation threshold.

\begin{figure}
\includegraphics[width=3.4in]{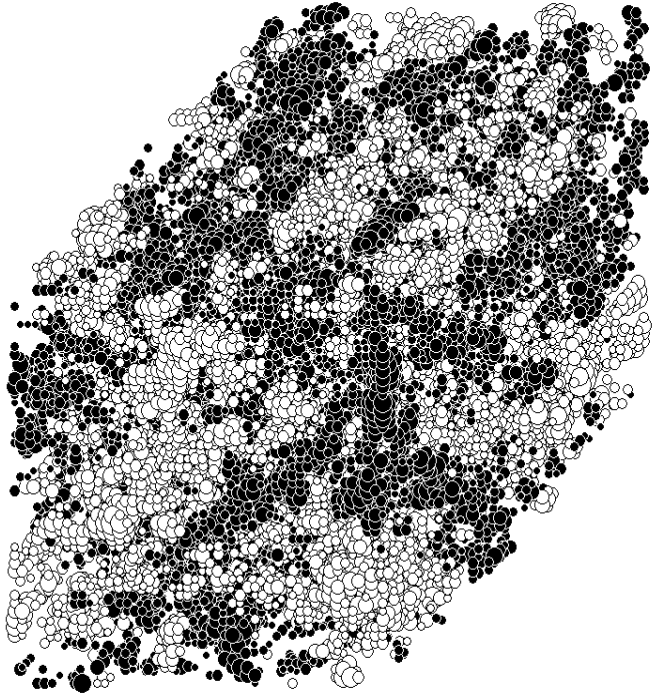}
\caption{Distribution of site energies $U$ in the lattice model of dipolar glass. A sample with the  size of $50\times 50\times 50$ lattice sites is shown. Black and white spheres represent the sites with positive and negative values of $U$, correspondingly, while the radius of a sphere is proportional to the absolute value of $U$. Sites with small absolute values of $|U|$ (less than $\sigma$) are not shown for the sake of clarity. }
\label{fig_clust}
\end{figure}

In this paper we are going to consider the statistics of $n_s$ for another case of correlated random distributions, namely for a Gaussian random field $U(\vec{r})$ (we will call $U(\vec{r})$ the random energy for reasons that will be obvious later), where the cluster may be defined as a connected set of sites, all of them having an energy greater than the threshold value $U_0 > 0$;  for a  Gaussian random field this is equivalent to the cluster with sites, having the energy less than $-U_0$. We are going to consider the distribution of "deep" clusters with $U_0\gg \sigma$ (where $\sigma$ is the rms disorder and where we assume zero average for $U$), i.e. the situation far away from the percolation threshold.  Most attention will be paid to the particular kind of a Gaussian random field having binary correlation function
\begin{equation}
C(\vec{r})=\left<U(\vec{r})U(0)\right> \approx A  \hskip2pt \sigma^2\frac{a}{r}, \hskip10pt r\gg a, \hskip10pt \sigma^2=\left<U^2(\vec{r})\right>,
\label{Cd}
\end{equation}
where the angular brackets denote a statistical averaging and $a$ is the lattice scale. This particular correlation function naturally arises in the model of dipolar glass (DG) \cite{Novikov:14573,Dunlap:542}, which is popular for the description of the charge transport properties of organic materials. In the simplest realization of the DG model we assume a random and independent orientation of dipoles occupying sites of a regular lattice, while charge carriers interact with dipoles by the long range charge-dipole interaction. In this model the energy of a charge carrier is
\begin{equation}
U(\vec{r})=e\sum_n \frac{\vec{d}_n\cdot(\vec{r}-\vec{r}_n)}{\varepsilon \left|\vec{r}-\vec{r}_n\right|^3},
\label{Ud}
\end{equation}
where $d$ is the dipole moment of the molecule, and $\varepsilon$ is the dielectric constant. Using an exact analytic calculation as well as computer simulations it was shown that for the DG model random energy $U(\vec{r})$ is a Gaussian random field if the average distance between dipoles is not significantly greater than the lattice scale \cite{Dieckmann:8136,Novikov:877e,Young:435}. The correlation function of the DG has the form (\ref{Cd}) and in the case of a simple cubic lattice $A\approx 0.76$ \cite{Dunlap:80}. This model was suggested to explain the Poole-Frenkel dependence of the carrier drift mobility $\mu$ in polar disordered organic materials on the applied electric field $E$ over a broad range of field strengths
\begin{equation}
\ln\mu\propto \sqrt{E}
\label{mu}
\end{equation}
 \cite{Dunlap:542,Novikov:130,Novikov:4472,Novikov:2584}. A power law decay of the correlation function (\ref{Cd}) means an extremely long range correlation in the random energy landscape in organic materials. For this reason  clusters have wide size distribution (see Fig. \ref{fig_clust}).

In this paper we show how to calculate analytically another characteristic of the correlated medium, namely, the distribution of the average random energy in a domain. An attractive feature of this distribution is that it is much more easy to calculate. We argue, then, that this distribution provides valuable information about the cluster numbers for large clusters and $U_0/\sigma \gg 1$, that is in the region far away from the percolation threshold.
 Our consideration will be limited to the 3D case, though generalization to other dimensions is obvious.

\section{Distribution of an average value of the random field in a domain}

Let us calculate the distribution $P_{V}(U_{0})$ of the average value $U_0$ of the random energy $U(\vec{r})$ in a domain with volume $V$ (here we consider a spatial average and use the same notation $U_0$ for the average energy). Let us start with a  continuous model of the medium, which is valid for $s\gg 1$. In this model the distribution $P_{V}(U_{0})$ is the average of the delta function
\begin{equation}
P_{V}(U_{0}) = \left<\delta\left(\frac{1}{V} \int{d\vec{r}U(\vec{r})f_V(\vec{r})} - U_{0}\right)\right>,
\label{def1}
\end{equation}
where $f_V(\vec{r})$ equals $1$ inside the domain and $0$ outside, and may be presented as a path integral over all realizations of the scalar field $U(\vec{r})$
\begin{eqnarray}
P_{V}(U_{0}) = \frac{1}{Z} \int{\mathcal{D}}U {\hskip
2pt}\delta\left(\frac{1}{V}\int{d\vec{r}U(\vec{r})f_V(\vec{r})} -
U_{0}\right)e^{-S},\\
Z = \int{\mathcal{D}}U e^{-S},\hskip10pt S = \frac{1}{2}\int{d\vec{r}d\vec{r}_1U(\vec{r})G(\vec{r}-\vec{r}_1)U(\vec{r}_1)}.\nonumber
\label{basic}
\end{eqnarray}
Here the kernel $G(\vec{r})$ obeys the equation
\begin{equation}
\int{d\vec{r}_2G(\vec{r}-\vec{r}_2)C(\vec{r}_2-\vec{r}_1)}=\delta (\vec{r}-\vec{r}_1).
\label{inv}
\end{equation}

To perform the actual integration we use the following presentation of the delta function
\begin{equation}
\delta(x) = \frac{1}{2\pi}\int{dye^{iyx}},
\label{delta}
\end{equation}
and then the Gaussian structure of the action $S$ allows us to calculate the integral (\ref{basic})
\begin{eqnarray}
\label{result}
P_{V}(U_{0}) = \frac{V}{\sqrt{2\pi K}}\exp \left(-\frac{U_0^2 V^2}{2K}\right),\\
K = \int{d\vec{r}d\vec{r}_1f_V(\vec{r})C(\vec{r}-\vec{r}_1)f_V(\vec{r}_1)}.\nonumber
\end{eqnarray}
This exact result is valid for any Gaussian field $U$. By definition, $C(0)=\left<U^2\right>$ and in a typical case $C(\vec{r})=\sigma^2 f(\vec{r})$. Hence, $K\propto \sigma^2$ and for this reason we introduce a new parameter $\kappa$
\begin{equation}
K=\kappa\sigma^2,
\label{k-space_k}
\end{equation}
which depends only on the spatial decay of $C(\vec{r})$. In future we will omit the factor $\sigma^2$ in $C(\vec{r})$. Using the Fourier transforms of $C(\vec{r})$ and $f_V(\vec{r})$ we obtain
\begin{equation}
\kappa = \frac{1}{(2\pi)^3} \int{d\vec{k}f_V(\vec{k})C(\vec{k})f_V(-\vec{k})},
\label{k-space1}
\end{equation}
and for a spherical domain with radius $R_0$
\begin{equation}
f_V(\vec{k})=\frac{4\pi}{k^3}\left(\sin
kR_0-kR_0\cos kR_0\right).
\label{k-space2}
\end{equation}

\subsection{Noncorrelated field}

 Let us analyze Eq. (\ref{result}) for some particular cases. If $U(\vec{r})$ is a field without spatial correlations, then
\begin{equation}
C(\vec{r}) =  a^3 \delta (\vec{r}),
\label{uncorr}
\end{equation}
and
\begin{equation}
\kappa= a^3 V, {\hskip 10pt} P_{V}(U_{0}) =
\left(\frac{V}{2\pi \sigma^2 a^3}\right)^{1/2}\exp \left(-\frac{U_0^2 V}{2\sigma^2
a^3}\right).
\label{ucK}
\end{equation}
Note, that in this particular case $P_{V}(U_{0})$ depends only on
the volume $V$ of the domain and not on its shape, as it should be
for a totally noncorrelated field distribution. Note also that
$V/a^3$ is actually the number $s$ of lattice sites in the
domain, so the leading asymptotics for $s\gg 1$ is
\begin{equation}
\ln P_{s} \propto - s.
\label{ucN}
\end{equation}

The noncorrelated Gaussian random energy is the base of the famous Gaussian disorder model, developed by H.~B{\"a}ssler for the description of charge carrier transport in disordered organic materials \cite{Bassler:15}. The correlated model \cite{Novikov:14573,Dunlap:542} could be considered as a natural extension of the B{\"a}ssler's model in order to explain the experimental mobility field dependence Eq. (\ref{mu}).

\subsection{Dipolar-like field in a spherical domain}

\begin{figure}
\includegraphics[width=3.4in]{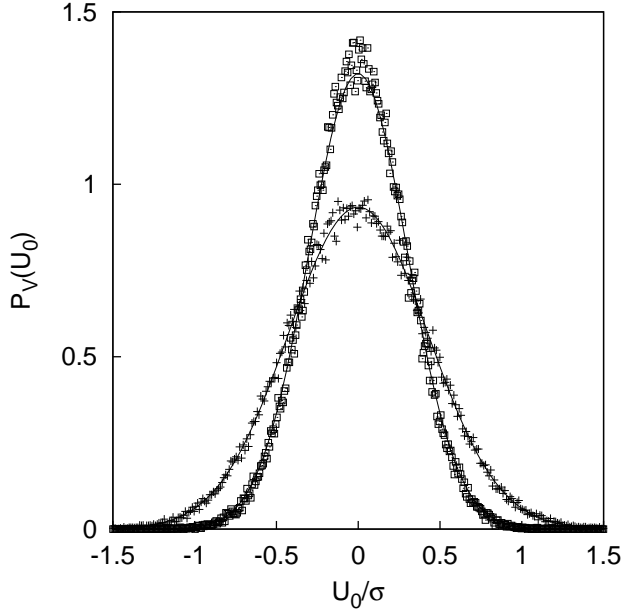}
\caption{Comparison between simulation data (+,$\Box$) and analytic results (\ref{result}) and (\ref{kappa_s})  for the dipolar glass with $R_0=5a$ (+) and $R_0=10a$ ($\Box$), respectively. Note that there are no adjustable parameters in this plot.}
\label{histogramm}
\end{figure}

\begin{figure}
\includegraphics[width=3.4in]{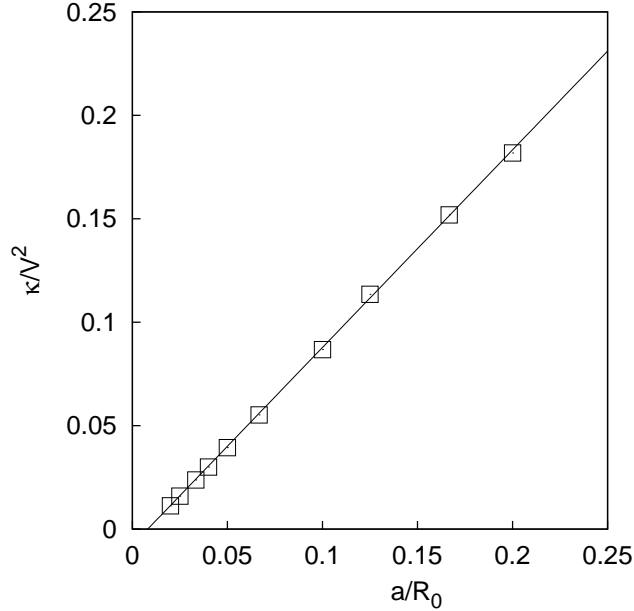}
\caption{Dependence of $\kappa$ on $R_0$ for a dipolar glass. The slope of the straight line equals to 0.95. According to Eq. (\ref{kappa_s}), it should be equal to 0.91. The reason for a nonzero intercept is the finite size of the basic cell.}
\label{s2R0}
\end{figure}

Now let us discuss the most interesting case of the dipolar-like
correlated field $C(\vec{r}) \propto a/r$. In this case $\kappa$ depends not only on the total volume of the domain, but also on its geometry. Using the Fourier transforms of $C(\vec{r})$
\begin{equation}\label{Cdip}
 C(\vec{k}) = \frac{4\pi A a}{k^2}
\end{equation}
we obtain for a spherical domain
\begin{equation}
\kappa = \frac{32\pi^2}{15}Aa R_0^5.
\label{kappa_s}
\end{equation}
This result demonstrates a tremendous difference with
the noncorrelated case, described by Eq. (\ref{ucK}), because in the leading asymptotics $\ln P_{V} \propto -R_0$, i.e. it is proportional to the linear size of the domain, and not to its volume. The comparison between the analytic result (\ref{kappa_s}) and the simulation data is shown in Figs. \ref{histogramm} and \ref{s2R0}. Statistics for all figures have been gathered for a  basic sample with a size of $256\times 256\times 256$ lattice sites with periodic boundary conditions and 10,000 realizations of the random field $U$ (apart from Fig. \ref{histogramm}, where 1,000 realizations of the random field were used). Particular distributions of $U(\vec{r})$ have been generated in the usual way. There  is  no  correlation  among  the  fluctuations in momentum space $U(\vec{k})$ for different $\vec{k}$, so we generated distributions of $U(\vec{k})$ and then calculated Fourier transform to get $U(\vec{r})$.

If a domain has an arbitrary shape but still could be characterized by a single linear scale $R_0$, then
\begin{equation}
\kappa \propto R_0^5
\label{prop}
\end{equation}
just because of dimensionality argument, though the
coefficient of proportionality depends on the actual shape of the domain. One can rewrite the relation (\ref{prop}) in the following form
\begin{equation}
\kappa = gAaV^{5/3},
\label{VS}
\end{equation}
where the coefficient $g$ depends on the shape of the domain and for a sphere $g_0=2(36\pi)^{1/3}/5$. The calculation of the coefficient $g$ for a more general case of elliptic domains is presented in the Appendix. This calculation shows that $g$ attains a maximum $g=g_0$ for a spherical shape and is significantly smaller than $g_0$ only for very elongated or oblate ellipsoids.

\section{Estimation for cluster numbers}

\begin{figure}
\includegraphics[width=3.4in]{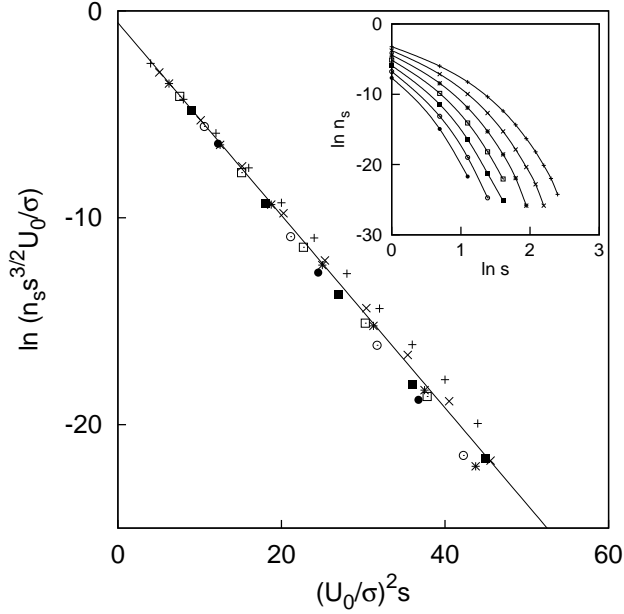}
\caption{Cluster numbers $n_s$ for a noncorrelated Gaussian field. Threshold energy $U_0/\sigma$ varies from 2.0 to 3.5 (with the step 0.25) from the topmost curve downwards, and the lines are provided as guides for an eye (inset). In proper coordinates all curves approximately collapse to a uniform straight line with the slope $-0.47$. According to Eq. (\ref{ucK}), the slope should be equal to $-1/2$.} \label{Nclust0_nc}
\end{figure}

\begin{figure}
\includegraphics[width=3.4in]{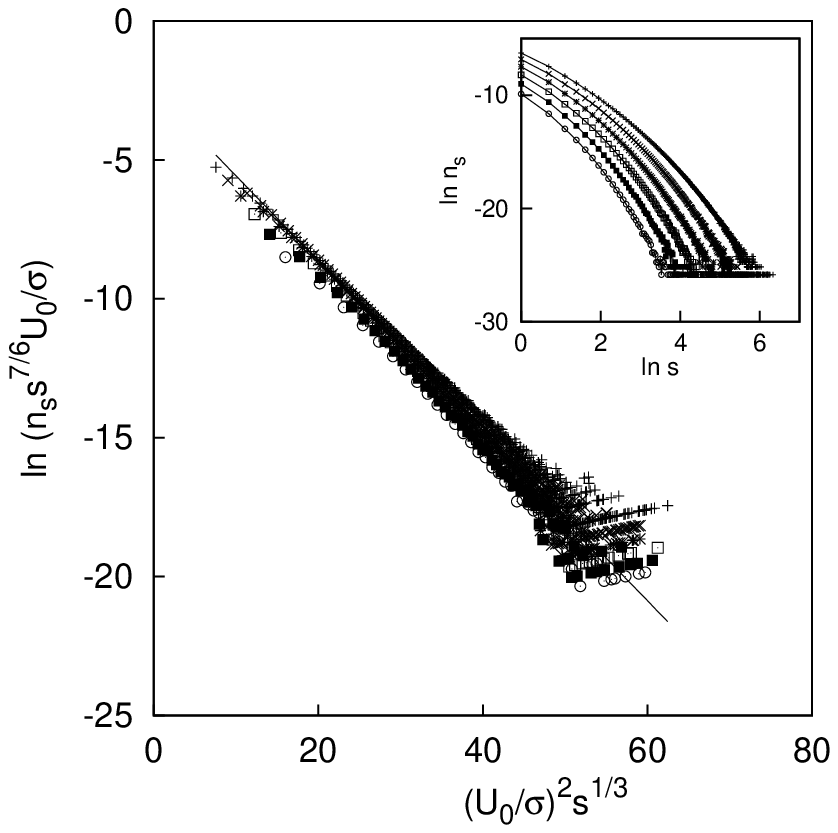}
\caption{Cluster numbers $n_s$ for the dipolar-like Gaussian field. Threshold energy
$U_0/\sigma$ varies from 2.75 to 4.0 (with the step 0.25) from the topmost curve downwards, and the lines are provided as guides for an eye (inset). Again, as in Fig. \ref{Nclust0_nc}, in proper coordinates all curves approximately collapse to the uniform straight line with the slope $-0.31$. According to Eq. (\ref{kappa_s}), the slope should be equal to $-0.34.$} \label{Nclust0}
\end{figure}

The number of spherical domains $n_V(U_0)$ per unit volume, having an average energy greater than $U_0$, is approximately equal to
\begin{equation}
n_V(U_0) \approx \frac{1}{V}\int_{U_0}^\infty dU P_V(U) =
\frac{1}{2V}\hskip2pt {\rm erfc}\left(\frac{U_0 V}{\sigma
\sqrt{2\kappa}}\right),
\label{erfc}
\end{equation}
here the coefficient $1/V$ reflects the number of non-overlapping independent domains in any finite sample. If $U_0 \gg \sigma$, then
\begin{equation}
n_V(U_0) \approx \frac{\sigma \sqrt{2\kappa}}{U_0 V^2
\sqrt{\pi}}\hskip2pt \exp\left(-\frac{U_0^2 V^2}{2\kappa
\sigma^2}\right).
\label{erfc3}
\end{equation}
We may expect that Eq. (\ref{erfc3}) gives a reasonable estimation for the number  $n_s$ of the true clusters, i.e. domains, where $U(\vec{r}) > U_0$ everywhere (assuming $V=a^3 s$), at least for the leading term of the asymptotic dependence of $n_s$ on $s$ (the very use of the continuous model of the random medium suggests that our consideration is valid only for $s\gg 1$). In addition, because we consider the distribution of the average field in the most compact domain (a sphere), this estimation could be valid only for clusters far away from the percolation threshold (this is equivalent to $U_0\gg\sigma$). At the percolation threshold clusters typically have a fractal-like structure \cite{Stauffer:book}. If this assumption is true, then for the noncorrelated Gaussian field
\begin{equation}
n_s \propto \frac{\sigma}{U_0s^{3/2}}\exp\left(-B_{nc}\frac{U_0^2}{\sigma^2} s\right),
\label{erfc4nc}
\end{equation}
and for the dipolar-like Gaussian field
\begin{equation}
n_s \propto \frac{\sigma}{U_0s^{7/6}}\exp\left(-B_{d}\frac{U_0^2}{\sigma^2} s^{1/3}\right),
\label{erfc4}
\end{equation}
where we take into account the possibility that for true clusters the coefficients $B_{nc}$ and $B_d$ might differ from the corresponding values $B_{nc}^0$ and $B_d^0$, estimated from Eqs. (\ref{ucK}) and (\ref{kappa_s}) for spherical domains
\begin{equation}
B_{nc}^0=\frac{1}{2},
\label{Bnc}
\end{equation}
\begin{equation}
B_{d}^0=\frac{5}{4A(36\pi)^{1/3}}=0.34...
\label{Bd}
\end{equation}
One can reasonably assume that $B_d$ does not differ significantly from $B_d^0$ because the spherical domains are the most probable ones (see Appendix). We compared Eqs. (\ref{erfc4nc}) and (\ref{erfc4}) with the simulation data and found that they provide good approximations for the true cluster numbers (see Figs. \ref{Nclust0_nc} and \ref{Nclust0}).

In order to understand the true status of Eqs. (\ref{erfc4nc})-(\ref{Bd}), let us compare the result for the noncorrelated field with Eq. (\ref{perc_deep}), which is an exact result for the noncorrelated percolation. The noncorrelated Gaussian field problem is exactly equivalent to the classic percolation problem \cite{Broadbent:629} with
\begin{equation}
p=\frac{1}{2}{\rm erfc}\left(\frac{U_0}{\sigma
\sqrt{2}}\right)\approx \frac{\sigma \sqrt{2}}{U_0
\sqrt{\pi}}\hskip2pt \exp\left(-\frac{U_0^2 }{2
\sigma^2}\right), \hskip10pt U_0/\sigma \gg 1.
\label{p_nc}
\end{equation}
Comparing Eqs. (\ref{perc_deep}), (\ref{erfc4nc}), and (\ref{p_nc}) we see that our simple estimation (\ref{erfc4nc}) provides at least the right leading asymptotics for $n_s$
\begin{equation}
\ln n_s =  -\frac{U_0^2}{2\sigma^2} s+o\left(\frac{U_0^2}{\sigma^2} s\right), \hskip10pt \frac{U_0^2}{\sigma^2} s \gg 1
\label{nca}
\end{equation}
(note that $B_{nc}=B_{nc}^0=1/2$), so both the functional kind of the asymptotic dependence of $n_s$ on $s$ \textit{and} the coefficient of proportionality are true for the noncorrelated field. In fact, even the small difference between $1/2$ and the corresponding fitting coefficient in Fig. \ref{Nclust0_nc} could be perfectly well explained by the contribution of higher order terms in Eq. (\ref{nca}) for $U_0/\sigma \simeq 2-3$. If we fit only the data for sufficiently large values of the threshold energy, then the slope becomes even more close to $-1/2$.

If we consider the case of a dipolar-like field, than again the fit of the true simulated $n_s$ to Eq. (\ref{erfc4}) gives $B_d\approx 0.31$ which is very close to $B_d^0\approx 0.34$. Again, if we try to fit only data points for $U_0/\sigma\simeq 3.5-4$, then the agreement between $B_d$ and $B_d^0$ becomes better. We would like to put forward the hypothesis that for the DG model the asymptotic expansion
\begin{equation}
\ln n_s =  -B_{d}^0\frac{U_0^2}{\sigma^2} s^{1/3}+o\left(\frac{U_0^2}{\sigma^2} s^{1/3}\right), \hskip10pt \frac{U_0^2}{\sigma^2} s^{1/3} \gg 1
\label{da}
\end{equation}
is valid too. If so, we may suggest that the corresponding asymptotics for $n_s$ and $s\gg 1$, $U_0\gg \sigma$ is valid for any Gaussian field with $\kappa$ calculated by Eq. (\ref{k-space1}) \textit{for a spherical domain}. This strong hypothesis certainly should be tested more thoroughly, but, nonetheless, our simulation data provides important arguments in its favor. Another interesting question is how valid are power law corrections to the leading exponents in Eqs. (\ref{erfc4nc}) and (\ref{erfc4}). For the  noncorrelated percolation in Eq. (\ref{perc_deep}) in the 3D case the exact result is $\theta=3/2$ \cite{Parisi:871} and agrees with Eq. (\ref{erfc4nc}), though this agreement, quite possibly, is an accident.

If we consider Eq. (\ref{nca}), it is obvious that it is universal and does not depend on the particular structure of the lattice. This is not so, seemingly, for Eq. (\ref{da}), where the coefficient $B_d^0$ depends on parameter $A$, which, in turn, is different for different lattices (the particular value $A\approx 0.76$ is valid only for a simple cubic lattice \cite{Dunlap:80}). At the same time, we cannot expect this kind of dependence for $s\gg 1$, where the particular structure of the lattice should be unimportant. This seeming contradiction could be resolved if we recall that for the DG model the parameter $\sigma^2$ depends on the lattice too. In fact, the combination $A\sigma^2a$ is invariant
\begin{equation}
A\sigma^2a=\frac{4\pi e^2d^2c}{3\varepsilon^2},
\label{As2}
\end{equation}
where $c$ is the concentration of dipoles \cite{Dunlap:80}. Clearly, in such a case the correlation function (\ref{Cd}) does not depend on any microscopic characteristic of the random dipolar medium, while the combination $B_d^0/\sigma^2$ depends only on the lattice scale $a$ and not on the particular structure of the lattice.

It was found previously that in the correlating percolation problem many features of the percolation near the percolation threshold are  not, in fact, very sensitive to the correlation. For example, in some cases the percolation threshold is the same for correlated and noncorrelated problems \cite{Tuthill:6389}, and cluster numbers sometimes are the same as well \cite{Jan:L705}. From this point of view it is very interesting that the asymptotic behavior of $n_s$ for deep clusters differs significantly for correlated and noncorrelated Gaussian fields.

\section{CONCLUSION}
In this paper we discussed the distribution of an average value in a finite domain for different Gaussian random fields. We found that for very different types of Gaussian fields  (in terms of their spatial correlation properties) the distribution of the average energy could serve as a good estimation for the true cluster numbers per lattice site  for large "deep" clusters, where $s\gg 1$ and the threshold energy $U_0$ is significantly greater than the rms disorder $\sigma$. Comparison of the analytical results for $P_V(U_0)$, calculated for a spherical domain, and computer simulation data for $n_s$ supports the hypothesis that $P_V(U_0)$ provides the exact leading asymptotic term for $n_s$. In our consideration we discussed particular Gaussian fields, relevant to the description of charge carrier transport in disordered organic materials. Nonetheless, the suggested approach could be used for other random Gaussian fields as well. Generalization to other spatial dimensions (beyond 3D) is also possible.

\section{ACKNOWLEDGMENTS}
Support from the ISTC (Grant No. 3718), the RFBR (Grants
No. 05-03-90579-NNS-a and 08-03-00125-a), the FWO, the Flemish Ministry of Education (Grant  No.  GOA 2006/2  and  Bilateral
Collaboration  Program,  BIL  05/34  Rusland), the Federal Science Policy of Belgium (Grant  No.  IAP-VI-27), are acknowledged.

\appendix*
\section{Dipolar-like field in an ellipsoidal domain}

\begin{figure}
\includegraphics[width=3.4in]{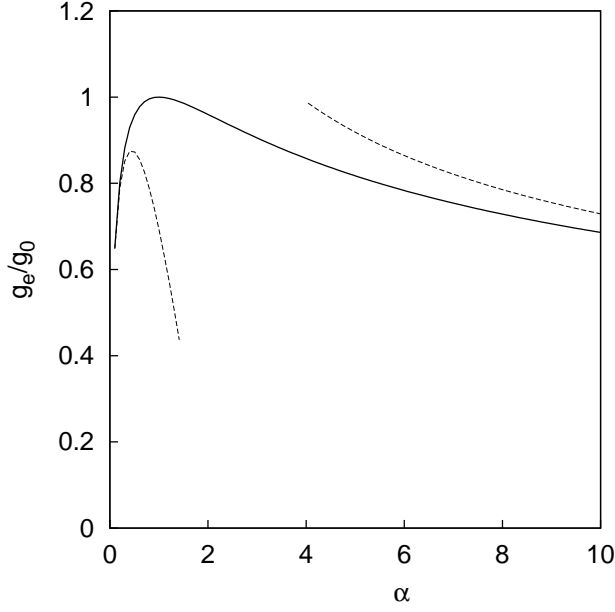}
\caption{Coefficient $g_e(\alpha)$ for the ellipsoidal domains; $g_0=2(36\pi)^{1/3}/5$ is the corresponding coefficient for a sphere. Broken lines correspond to approximations (\ref{cigar_f}) and (\ref{pan_f}). }
\label{g(a)}
\end{figure}

Let us consider domains of a non-spherical shape in order to estimate the influence of the domain shape on the probability to have a particular value of $U_0$ in such domain. The simplest choice is to study the distribution of the field in ellipsoidal domains with half-axes $b_1R_0$, $b_2R_0$, and $b_3R_0$, where $b_i$ are scale coefficients. A direct calculation shows that for ellipsoidal domains
\begin{eqnarray}
\kappa=\kappa_e = \frac{b_1 b_2 b_3}{(2\pi)^3} \int{d\vec{k}f_{V_0}(\vec{k})C_e(\vec{k})f_{V_0}(-\vec{k})}, \\
C_e(\vec{k}) = \frac{4\pi Aa}{\sum_{i=1}^{3}k^2_i/b^2_i},{\hskip 30pt}\nonumber
\label{ellips1}
\end{eqnarray}
here the function $f_{V_0}(\vec{k})$ is exactly the same one as the
corresponding function for spherical domains and $V_0=4\pi R_0^3/3$. For this reason
\begin{equation}
\kappa_e =  \frac{1}{2}\kappa_0 b_1b_2b_3I,
\end{equation}
\[
I=b_1 b_2\int_{-1}^{1}
\frac{dx}{\left\{\left[1+x^2\left(\frac{b_1^2}{b_3^2}-1\right)\right]
\left[1+x^2\left(\frac{b_2^2}{b_3^2}-1\right)\right]\right\}^{1/2}},\nonumber
\label{angle2}
\]
here $\kappa_0$ is the corresponding value for a sphere. Let us calculate this integral for ellipsoidal domains having rotational symmetry with $b_1=b_2$. In this case
\begin{eqnarray}
I= b_1^2\int_{-1}^{1}{\frac{dx}{1+
\left(\alpha^2-1\right)x^2}}= \\
=\frac{2 b^2_1}{\sqrt{|\alpha^2-1|}}\begin{cases}
\text{arctg}\sqrt{\alpha^2-1},& \alpha > 1,\\
\frac{1}{2}\ln\frac{1+\sqrt{1-\alpha^2}}{1-\sqrt{1-\alpha^2}},&
\alpha < 1,
\end{cases}\nonumber
\label{rot}
\end{eqnarray}
where $\alpha=b_1/b_3$. Taking into account that the volume of the ellipsoidal domain is equal to $V=b_1 b_2 b_3 V_0$, we obtain
\begin{equation}
\frac{g_e(\alpha)}{g_0}=\frac{\alpha^{2/3}}{\sqrt{|\alpha^2-1|}}\begin{cases}
\text{arctg}\sqrt{\alpha^2-1},& \alpha > 1,\\
\frac{1}{2}\ln\frac{1+\sqrt{1-\alpha^2}}{1-\sqrt{1-\alpha^2}},&
\alpha < 1,
\end{cases}
\label{gall}
\end{equation}
and in the limiting cases
\begin{equation}
\frac{g_e(\alpha)}{g_0}\approx \alpha^{2/3}\ln\frac{2}{\alpha}, {\hskip 10pt} \alpha \ll 1,
\label{cigar_f}
\end{equation}
\begin{equation}
\frac{g_e(\alpha)}{g_0}\approx  \frac{\pi}{2\alpha^{1/3}}, {\hskip 10pt} \alpha \gg 1.
\label{pan_f}
\end{equation}
Eqs. (\ref{cigar_f}) and (\ref{pan_f}) mean that domains which differ significantly from the spherical ones have much smaller probability to occur (for the same values of $U_0$ and $V$). The general behavior of $g_e(\alpha)$ is shown in Fig. \ref{g(a)}.

\end{document}